\documentclass[twocolumn,nofootinbib,amsmath,amssymb,a4paper]{revtex4}

\usepackage{graphicx}
\usepackage{dcolumn}
\usepackage{bm}

\newcommand{\derzero}{\includegraphics[scale=.3]{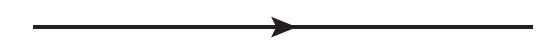}}
\newcommand{\derone}{\raisebox{-0.3\totalheight}{\includegraphics[scale=.3]{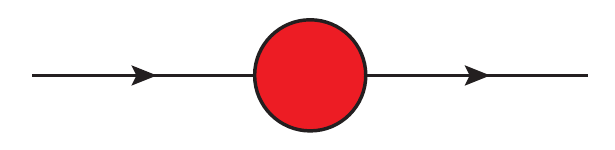}}}
\newcommand{\dertwoa}{\raisebox{-0.3\totalheight}{\includegraphics[scale=.3]{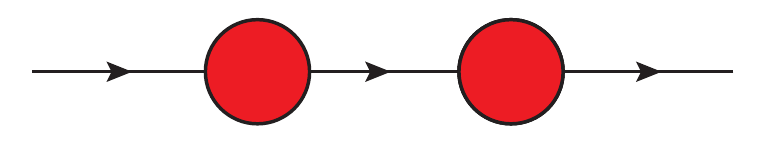}}}
\newcommand{\dertwob}{\raisebox{-0.3\totalheight}{\includegraphics[scale=.3]{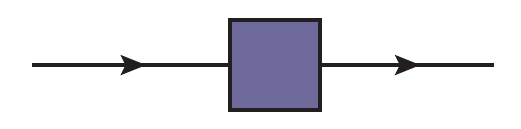}}}

\newcommand{\insV}{\raisebox{-0.3\totalheight}{\includegraphics[scale=.3]{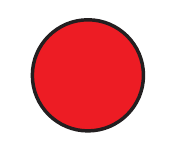}}}
\newcommand{\insT}{\raisebox{-0.3\totalheight}{\includegraphics[scale=.3]{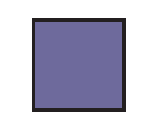}}}
\newcommand{\axialdiagram}{\vcenter{\hbox{\includegraphics[scale=.35]{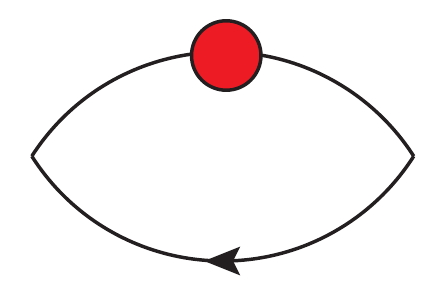}}}}

\newcommand{\ifp}{\raisebox{-0.08\totalheight}{\includegraphics[scale=.3]{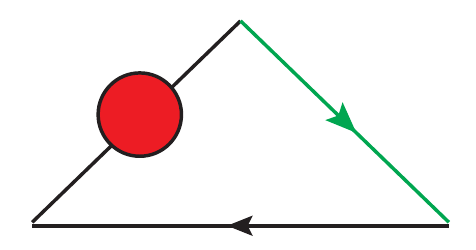}}}
\newcommand{\ifq}{\raisebox{-0.08\totalheight}{\includegraphics[scale=.3]{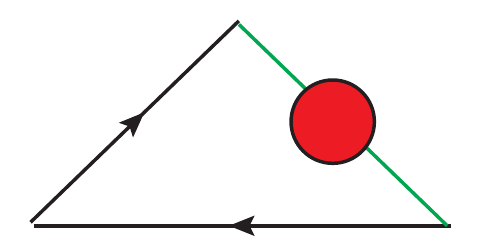}}}
\newcommand{\fip}{\raisebox{-0.08\totalheight}{\includegraphics[scale=.3]{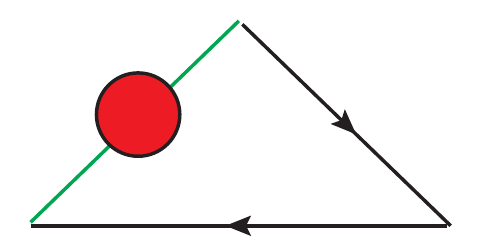}}}
\newcommand{\fiq}{\raisebox{-0.08\totalheight}{\includegraphics[scale=.3]{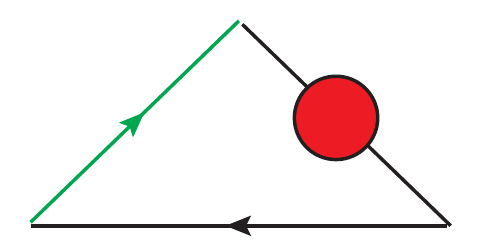}}}
\newcommand{\iiq}{\raisebox{-0.08\totalheight}{\includegraphics[scale=.3]{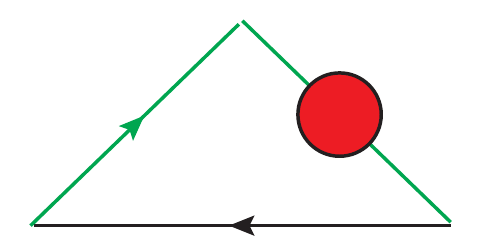}}}
\newcommand{\iip}{\raisebox{-0.08\totalheight}{\includegraphics[scale=.3]{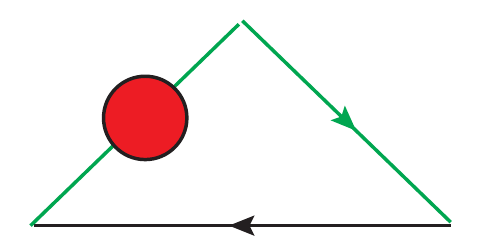}}}
\newcommand{\ffq}{\raisebox{-0.08\totalheight}{\includegraphics[scale=.3]{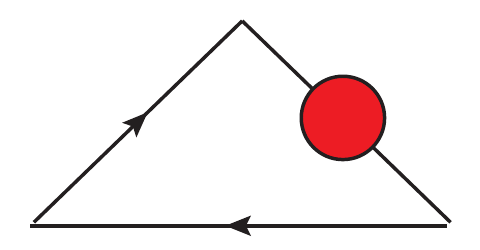}}}
\newcommand{\ffp}{\raisebox{-0.08\totalheight}{\includegraphics[scale=.3]{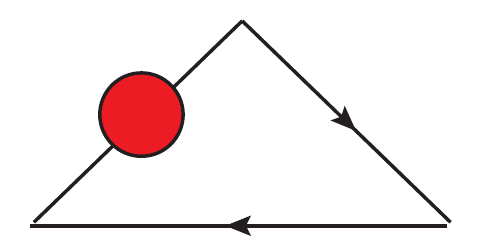}}}

\newcommand{\conn}{\vcenter{\hbox{\includegraphics[scale=.25]{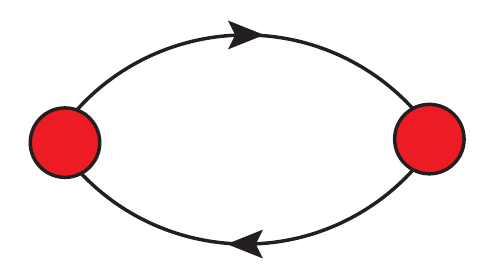}}}}
\newcommand{\disc}{\vcenter{\hbox{\includegraphics[scale=.3]{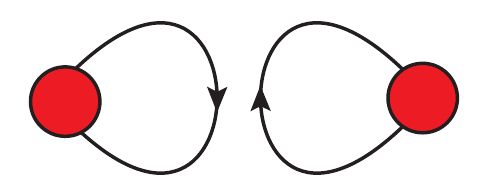}}}}

\newcommand{\Pa}{\vcenter{\hbox{\includegraphics[scale=.1]{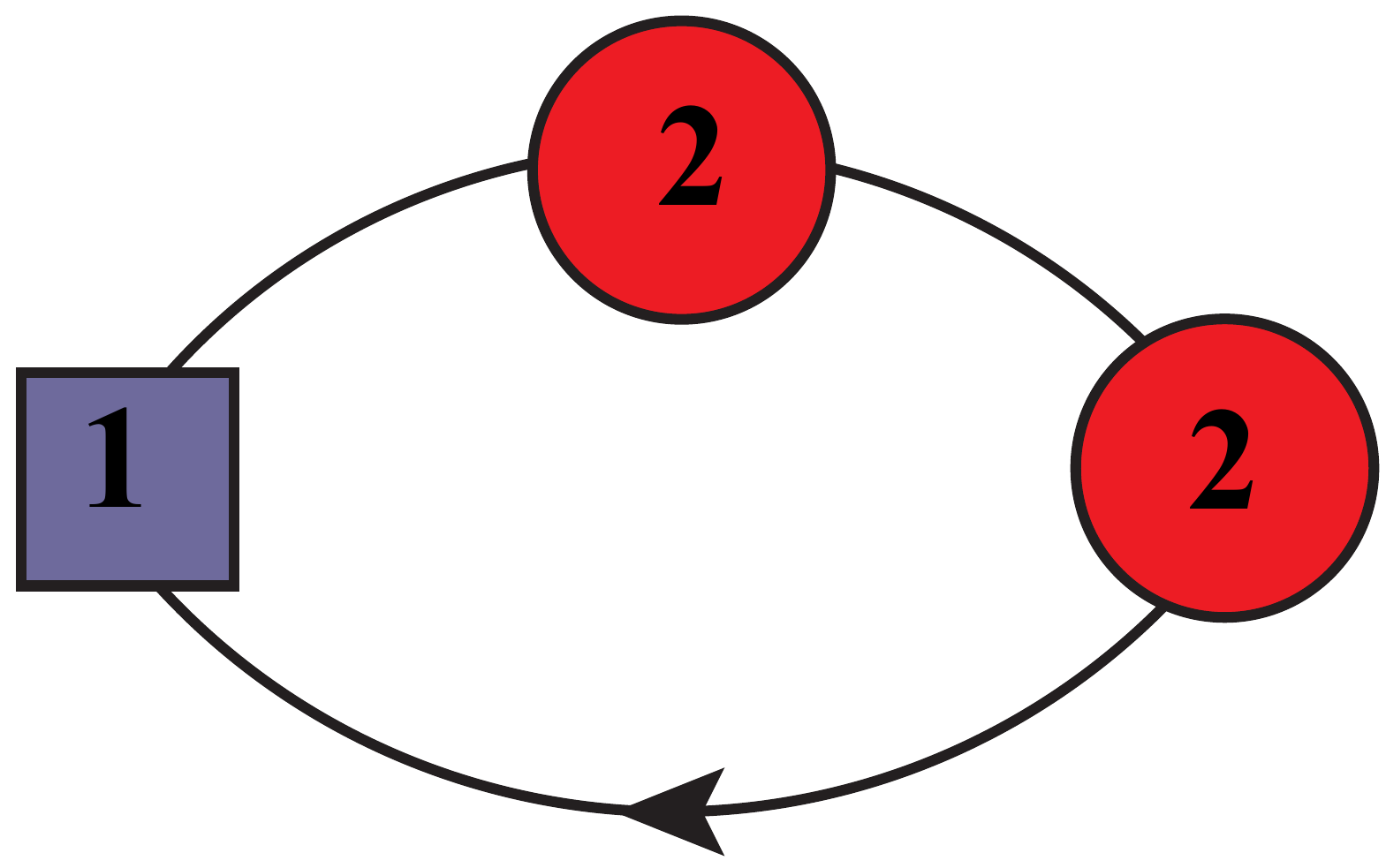}}}}
\newcommand{\Pb}{\vcenter{\hbox{\includegraphics[scale=.1]{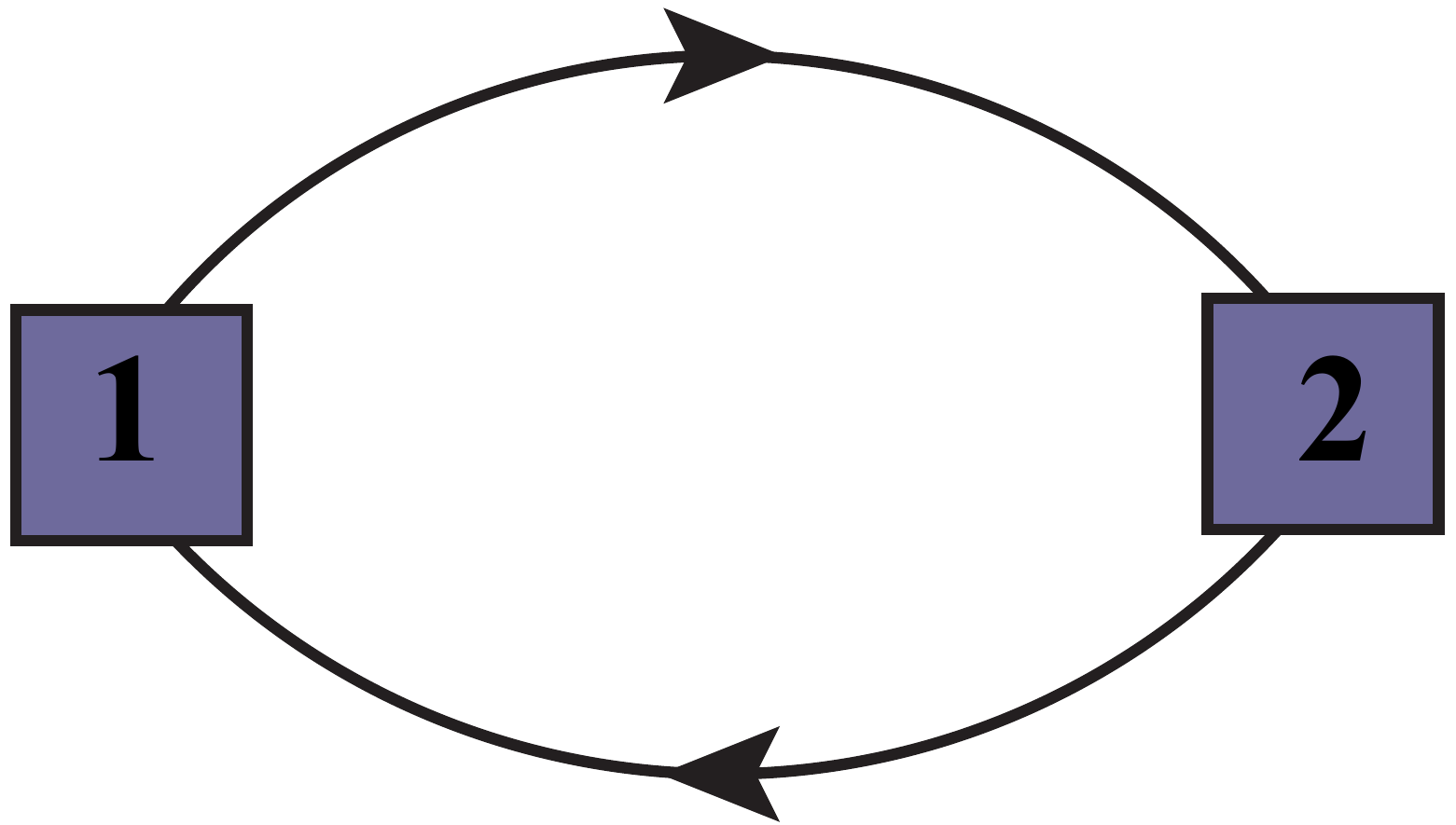}}}}
\newcommand{\Pcone}{\vcenter{\hbox{\includegraphics[scale=.1]{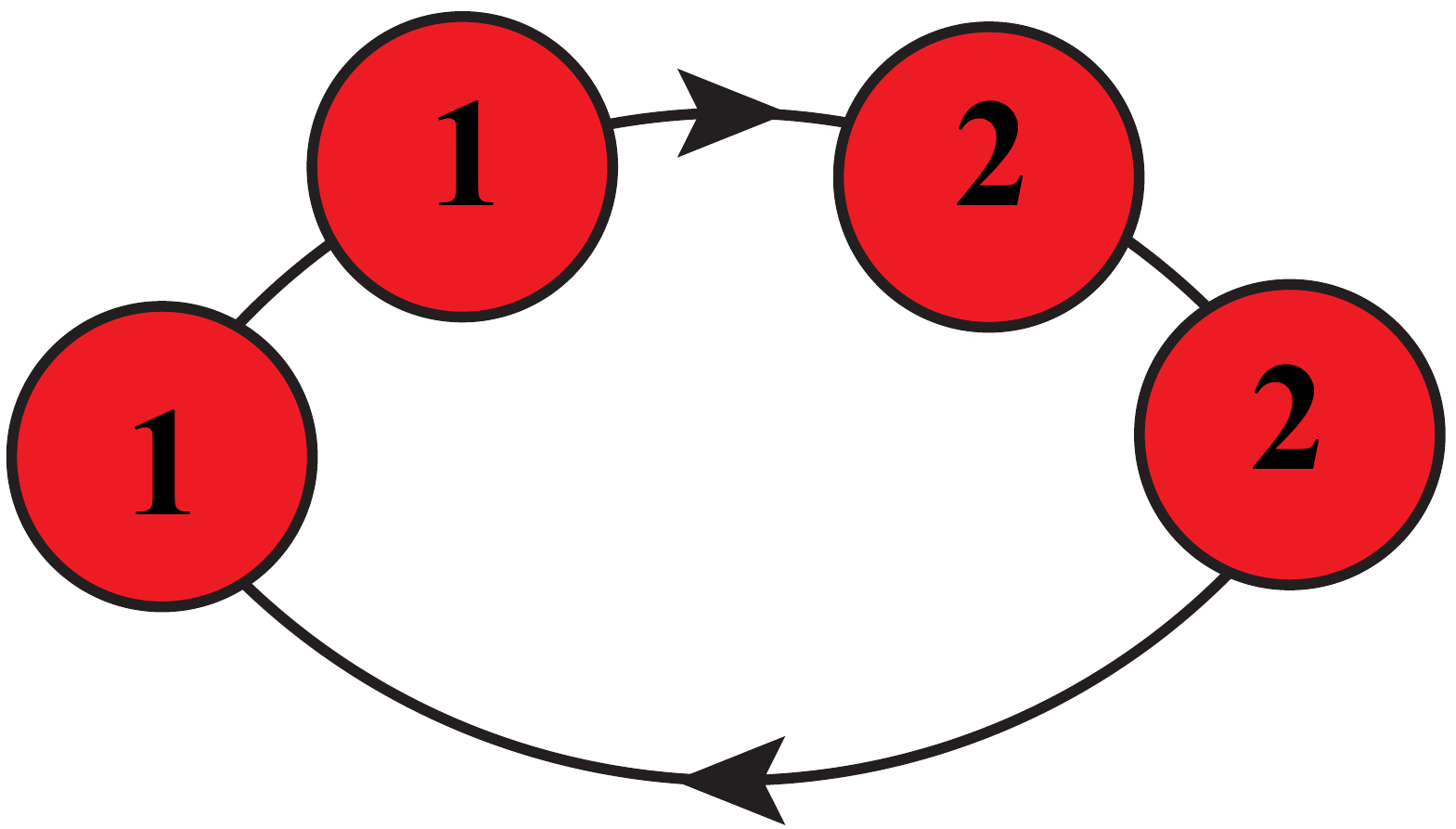}}}}
\newcommand{\Pctwo}{\vcenter{\hbox{\includegraphics[scale=.1]{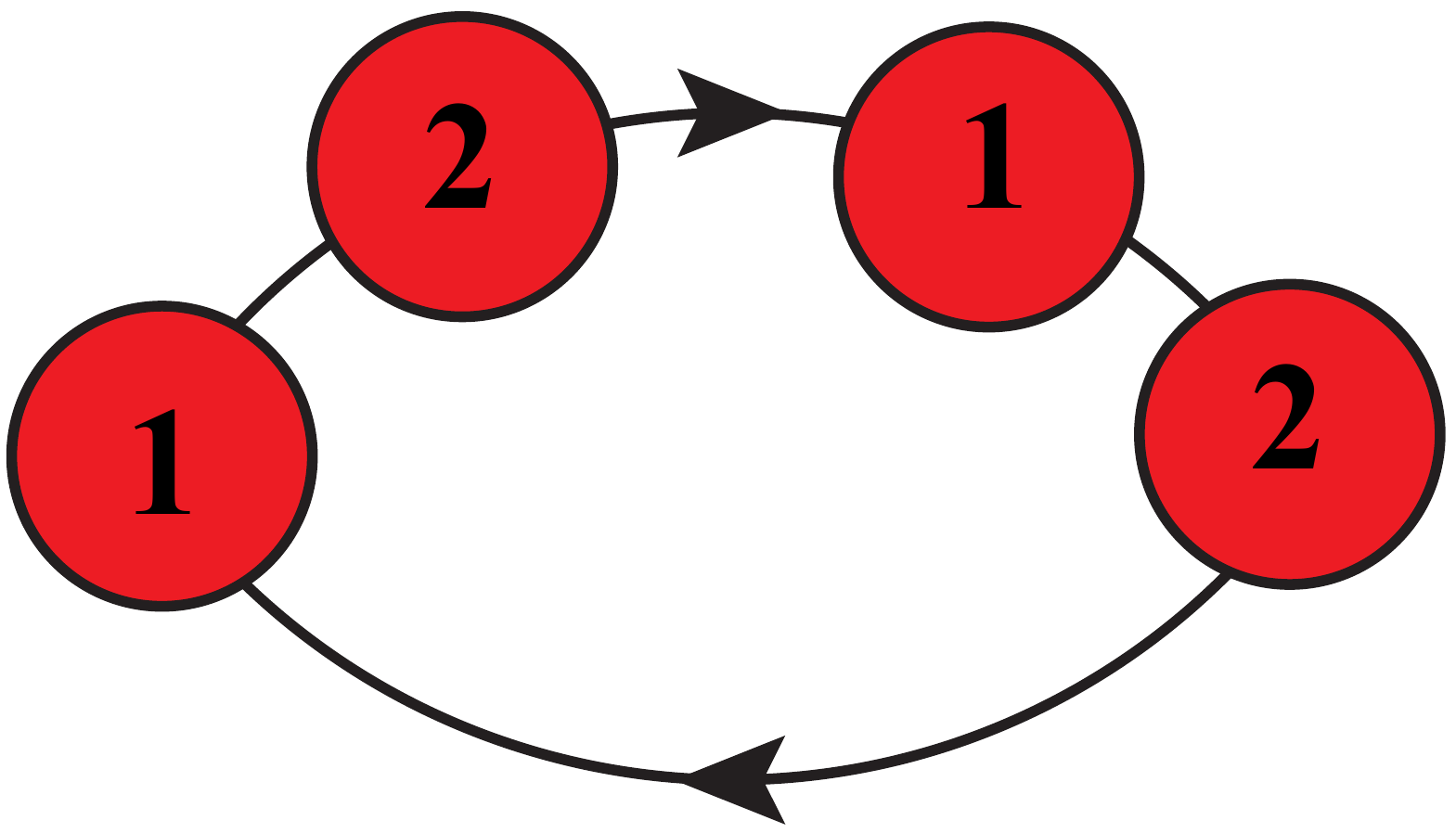}}}}
\newcommand{\Pd}{\vcenter{\hbox{\includegraphics[scale=.1]{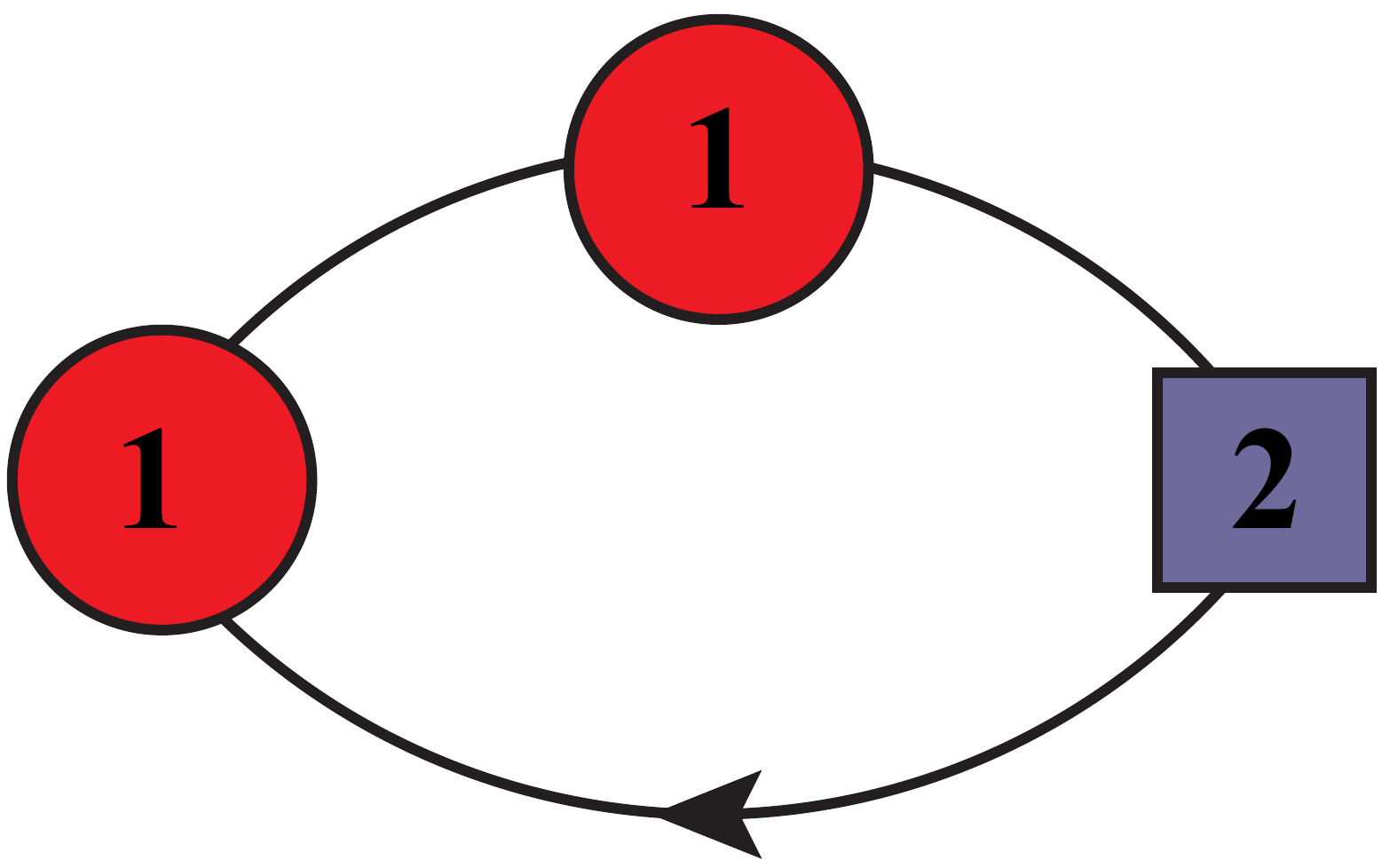}}}}

\begin{document}

\title{On the extraction of zero momentum form factors on the lattice}

\author{G.M. de Divitiis, R. Petronzio, N. Tantalo}
\affiliation{\vskip 10pt
Universit\`a di Roma ``Tor Vergata'', I-00133 Rome, Italy\\
INFN sezione di Roma ``Tor Vergata'', I-00133 Rome, Italy\\
}%

\begin{abstract}
We propose a method to expand correlation functions with respect to the spatial components of external momenta. From the coefficients of the expansion it is possible to extract Lorentz--invariant form factors at zero spatial momentum transfer avoiding model dependent extrapolations. These objects can be profitably calculated on the lattice. 
We have explicitly checked the validity of the proposed procedure by considering two-point correlators with insertions of the axial current, the form factors of the semileptonic decay of pseudoscalar mesons, and the hadronic vacuum polarization tensor entering, for example, the lattice calculation of the anomalous magnetic moment of the muon. 
\end{abstract}

\maketitle
\section{introduction}
Physical observables are calculated on the lattice by extracting hadronic matrix elements from the large euclidean time behavior of correlation functions. 
The matrix elements are Lorentz--covariant quantities, i.e. tensors of a certain rank and with definite properties under discrete symmetries. These are parametrized in terms of form factors, the Lorentz--invariant coefficients associated with the decomposition on a certain basis. The basis is built by considering products of external momenta, polarization vectors, metric tensor, etc. The form factors multiplying powers of the the external momenta are difficult to extract when these variables vanish.

The class of problems we address in this letter can be exemplified as follows
\begin{eqnarray}
C^{\mu\cdots}(p,\cdots)= p^\mu F^{\cdots}(p^2,\cdots) \;,
\end{eqnarray}
where $C^{\mu\cdots}$ is the quantity that can be computed directly on the lattice (a correlator or a combination of correlators), $F^{\cdots}$ is the quantity one is interested in, while the dots represent additional arguments that may or may not be present and that characterize the given observable. 

Typically it is possible to obtain good numerical precision on the form factors for values of $p^\mu$ in a limited range. Lattice correlators tend to be extremely noisy for large values of the modulus of $p^\mu$. On the other side, although the use of flavour twisted boundary conditions~\cite{Bedaque:2004kc,deDivitiis:2004kq} allows to overcome the quantization of spatial momenta on finite volumes and to compute $F^{\cdots}$ for very small values of $p^\mu$, the numerical signal for $C^{\mu\cdots}$ goes to zero for nearly vanishing values of $p^\mu$ and the results in this region are again very rough. In particular there are situations in which one is interested in the value of $F^{\cdots}$ at $p^\mu=0$ that, by using standard lattice techniques, can only be obtained by recurring to (model dependent) extrapolations. 

A concrete example of this kind is offered by the calculation at zero recoil of all the independent form factors parameterizing the semileptonic decays $B\rightarrow D^{(\star)}\ell\nu$~\cite{Hashimoto:1999yp,de Divitiis:2007ui,de Divitiis:2007uk,deDivitiis:2008df,Bernard:2008dn}.

Another phenomenologically relevant example is given by the hadronic vacuum polarization contribution to the muon $g-2$ that is extracted from the integrated two--point correlator of the electromagnetic currents of the quarks,
\begin{eqnarray}
C^{\mu\nu}(p)&=&
\frac{1}{(TL^3)^2}\sum_{x,y}e^{ip(y-x)} \langle 
V^\mu_{em}(x)\ V^\nu_{em}(y)
\rangle
\nonumber \\
\nonumber \\
&=&(\delta^{\mu\nu}p^2-p^\mu p^\nu)\Pi(p^2) \; .
\end{eqnarray}
In this case the accurate knowledge of $\Pi(0)$ is extremely important because divergent contributions are canceled by extracting the physical information from the difference $\Pi(p^2)-\Pi(0)$. By using standard approaches, $\Pi(0)$ can only be obtained by extrapolating the results at non vanishing $p^\mu$ and by relying on fitting formulae that unavoidably introduce systematic errors (see ref.~\cite{Aubin:2012me} for a recent discussion of this point).

In this work we discuss a method to be used in order to solve the class of problems discussed above. The method allows to calculate \emph{directly} on the lattice the $n$th-derivative of a lattice correlator with respect to external spatial momenta. By using the notation of the previous formulae, we shall describe in the following sections a procedure to calculate on the lattice the left hand side of (no sum over $k$)
\begin{eqnarray}
\left . \frac{\partial C^{k\cdots}(p,\cdots)}{\partial p^k}\right\vert_{p^k=0}
= \left. F^{\cdots}(p^2,\cdots)\right\vert_{p^k=0} \; ,
\end{eqnarray}
directly at $p^k=0$ without the need of any extrapolation and related ansatz for the fitting function. 
The same problem was also addressed in refs.~\cite{Aglietti:1994nx,Lellouch:1994zu} with a solution different from the one proposed 
in this paper.

By applying this method it is possible to compute on the lattice the vacuum polarization scalar form factor at zero momentum according to
\begin{eqnarray}
\Pi(0)=-\left. \frac{\partial^2C^{12}(p)}{\partial p^1 \partial p^2} \right\vert_{p^2=0} \; .
\end{eqnarray}
%

\section{Expansion of the correlators}
In order to expand lattice observables in powers of external spatial momenta,
let us start by considering a generic correlator depending upon $\vec p$,
\begin{eqnarray}
C(\vec p) 
&=&
\frac{1}{L^3}\sum_{\vec x} e^{-i\vec p \cdot \vec x} \int dU \, P[U]\, C[\vec x;U] 
\nonumber \\
\nonumber \\
&=&
\int dU \,P[U]\, C[\vec p;U] \; .
\end{eqnarray}
In previous expressions $P[U]$ represents the gauge ``probability density'' and incorporates  the fermionic determinants together with the normalization factor of the functional integral. Note that $P[U]$ does not depend upon external momenta. Our method consists in expanding $C[\vec p;U]$, i.e. the correlator at fixed QCD gauge background, with respect to $\vec p$ according to
\begin{eqnarray}
C[\vec p;U]= C^{(0)}[U]+p_k C^{(1)}_k[U]+\frac{p_hp_k}{2}C^{(2)}_{hk}[U]+\cdots \; .
\nonumber \\
\end{eqnarray}
The coefficients $C^{(n)}_{\cdots}[U]$ can be interpreted as the derivatives of the original correlator with respect to the external momenta. Note that this is true also when the external momenta are quantized as a consequence of the finite volume and of the boundary conditions satisfied by the quark and gauge fields.

After fermionic integration and Wick contractions the lattice correlator $C[\vec p;U]$ can be expressed as the product of quark propagators and operator insertions, i.e.
\begin{eqnarray}
C(\vec p) 
&=&
\int dU P[U] \frac{1}{L^3}\sum_{\vec x,\cdots} e^{i\vec p \cdot (\vec y-\vec x)}
\mbox{Tr}\left\{ S[x,y;U]\, \Gamma \cdots \right. 
\nonumber \\
\nonumber \\
&=&
\langle\frac{1}{L^3}\sum_{\vec x,\cdots}
\mbox{Tr}\left\{ S[x,y;U,\lambda^p]\, \Gamma \cdots \right. \rangle \;.
\nonumber \\
\end{eqnarray}
Here $\Gamma$ represent a generic operator insertion (see section~\ref{sec:expvert} below)
while $S[x,y;U]$ is one of the fermion propagators entering the expression of the given correlator. The fermion propagator $S[x,y;U]$ ($S[U]$ in short) satisfies the following equation
\begin{eqnarray}
\sum_y D[x,y;U]\ S[y,z;U] = \delta_{x,z} \; ,
\end{eqnarray}
where $D[x,y;U]$ is the lattice Dirac operator used to discretize the quarks action. 
The propagator $S[x,y;U,\lambda^p]$ introduced above ($S[U,\lambda^p]$ in short) is simply given by
\begin{equation}
S[x,y;U,\lambda^p]  = e^{i\vec p \cdot (\vec y - \vec x)} S[x,y;U] \; .
\end{equation}
The previous definition has been made because it is useful to obtain $S[U,\lambda^p]$ directly, i.e. by solving the lattice Dirac equation with  the link variables rescaled by a phase factor
\begin{eqnarray}
&&U_k(x) \ \longmapsto \ \lambda_k^p U_k(x) = e^{ip_k}\, U_k(x) \; ,
\nonumber \\ 
\nonumber \\
&&\sum_y D[x,y; U, \lambda^p]\ S[y,z;U, \lambda^p] = \delta_{x,z}  \; .
\end{eqnarray}
Note that $S[U,\lambda^p]$ is defined analogously to a lattice propagator associated with a quark field satisfying flavour twisted boundary conditions~\cite{Bedaque:2004kc,deDivitiis:2004kq},
with the difference that now $\vec p$ is not a continuous variable. 
In this work we shall use periodic boundary conditions for all the quark fields and, consequently, we shall always work within a unitary setup and with the spatial momenta quantized according to
\begin{eqnarray}
\vec p = \frac{2\pi}{L}\left(n_1,n_2,n_3\right) \;.
\end{eqnarray}
Expressions for the coefficients $C^{(n)}_{\cdots}[U]$ can now be readily obtained by
expanding the lattice propagators $S[U,\lambda^p]$ and possibly non ultra-local operators entering into the original expression of $C[\vec p;U]$ (see the following sections for examples) in powers of the external momenta. Indeed, since the quark determinants do not depend upon external momenta, the derivatives of the correlator do not involve additional disconnected fermionic Wick contractions with respect to the ones originally present to define $C[\vec p;U]$.
\section{Expansion of the propagators}
The expansion of the quark propagators can be obtained by starting from the expansion of the lattice Dirac operator,
\begin{eqnarray}
&&\lambda_k^p=1+i {p_k }-\frac{1}{2} p_k^2+\cdots \;,
\nonumber \\
\nonumber \\
&&D[U,\lambda^p]  = 
D[U]+
p_k \left. \frac{\partial D}{\partial p_k} \right\vert_{\vec p=0}+ 
\frac{p_k^2}{2} \left. \frac{\partial^2D}{\partial p_k^2}\right\vert_{\vec p=0}+\cdots \;,
\nonumber \\
\end{eqnarray}
and by using the identity $D[U,\lambda^p] S[U,\lambda^p]=1$ leading to
\begin{eqnarray}
&&\frac{\partial S}{\partial p_k} = -S \frac{\partial D}{\partial p_k} S \; ,
\nonumber \\
\nonumber \\
&&\frac{1}{2}\frac{\partial^2S}{\partial p_k^2} =
+ \,S \frac{\partial D}{\partial p_k} S \frac{\partial D}{\partial p_k} S
-S \frac{1}{2} \frac{\partial^2D}{\partial p_k^2} S \; ,
\nonumber \\
\nonumber \\
&&\cdots  \; .
\label{eq:textpropexp}
\end{eqnarray}
In the previous expressions there is no sum over $k$ and we have used the following compact notation
\begin{eqnarray}
\frac{\partial \mathcal{O}}{\partial p_k}\equiv 
\left . \frac{\partial \mathcal{O}[\cdots;U,\lambda^p]}{\partial p_k} \right\vert_{\vec p=0} \; .
\end{eqnarray}

For our numerical calculations we have used clover Wilson fermions, corresponding to
the specific Dirac operator $D[U,\lambda^p]$ given below
\begin{eqnarray}
&&\sum_{x,y} \bar \psi(x) D[x,y; U, \lambda^p]  \psi(y) = 
\nonumber \\
\nonumber \\
\nonumber \\
&&\sum_{x}{\bar \psi(x) (4+m)  \psi(x)}
+\frac{ic_{SW}}{4}\sum_{x,\mu \nu}{ 
\bar \psi(x) \sigma_{\mu \nu} F^{\mu \nu}(x)\, \psi(x)}
\nonumber \\
\nonumber \\
&&\qquad\qquad\qquad -
\sum_{x,\mu} {\bar \psi(x)\lambda_\mu^p U_\mu(x)\frac{1-\gamma^\mu}{2}\psi(x+\mu)}
\nonumber \\
\nonumber \\
&&\qquad\qquad\qquad -
\sum_{x,\mu} {\bar \psi(x+\mu) (\lambda_\mu^p)^\dagger U_\mu^\dagger(x) \frac{1+\gamma^\mu}{2}\psi(x)} \; ,
\nonumber \\
\end{eqnarray}
where $\lambda_0^p = 1$, $\lambda_k^p = e^{ip_k}$, $F^{\mu \nu}(x)$ is the clover definition of the lattice gluon field strength tensor and $c_{SW}$ is the Sheikholeslami--Wohlert coefficient~\cite{Sheikholeslami:1985ij}.
The $n$th-derivatives of this operator are proportional either to the ``point--split" vector current $V^k(x,\vec 0)$ (for odd $n$), or to the ``tadpole" current $T^k(x,\vec 0)$ (for even $n$),
\begin{eqnarray}
&&\left[ \bar \psi \frac{\partial D}{\partial p_k}  \psi\right](x) = i \, V^k(x,\vec 0) 
\nonumber \\ 
\nonumber \\
&&\left[\bar \psi \frac{\partial^2 D}{\partial p_k^2} \psi\right](x) = T^k(x,\vec 0)  
\nonumber \\ 
\nonumber \\
&&\left[\bar \psi \frac{\partial^n D}{\partial p_k^n} \psi\right](x) = - (- i)^n 
\left\{
\begin{array}{c}
V^k (x,\vec 0) \text{ odd } n
\\
\\
T^k (x,\vec 0) \text{ even } n
\end{array}
\right. \; ,
\nonumber \\
\end{eqnarray}
where the expressions of the two currents are given by
\begin{eqnarray}
V^\mu(x,\vec p) &=& \left[ \bar \psi \Gamma_V^\mu \psi\right](x,\vec p)
\nonumber \\
\nonumber \\
&=&+
\bar \psi(x+\mu) (\lambda_\mu^p)^\dagger U_\mu^\dagger(x) \frac{1+\gamma^\mu}{2}\psi(x)
\nonumber \\
\nonumber \\
&&-
\bar \psi(x)\lambda_\mu^p U_\mu(x) \frac{1-\gamma^\mu}{2}\psi(x+\mu) \; ,
\nonumber \\
\nonumber \\
\nonumber \\
T^\mu(x,\vec p)&=& \left[ \bar \psi \Gamma_T^\mu \psi\right](x,\vec p)
\nonumber \\
\nonumber \\
&=&+
\bar \psi(x+\mu) (\lambda_\mu^p)^\dagger U_\mu^\dagger(x) \frac{1+\gamma^\mu}{2}\psi(x)
\nonumber \\
\nonumber \\
&&+
\bar \psi(x)\lambda_\mu^p U_\mu(x) \frac{1-\gamma^\mu}{2}\psi(x+\mu) \; .
\nonumber \\
\end{eqnarray}
Graphically, the expansion of the propagator of eqs.~(\ref{eq:textpropexp}) can be represented  as the sum of iterated insertions at zero momentum of the operators 
$V^k (x,\vec 0)$ and $T^k (x,\vec 0)$ according to (no sum over $k$) 
\begin{eqnarray}
&&\overset{p_k} {\derzero}=
\derzero-i p_k \derone
\nonumber \\
\nonumber \\
&&\qquad +
( i p_k)^2 \left \{\dertwoa + \frac{1}{2} \dertwob \right\} +
\cdots\; .
\nonumber \\
\label{eq:graphvertexp}
\end{eqnarray}
In the previous expressions we have introduced the following graphical notation
\begin{eqnarray}
_y\derzero_x &=& S[x,y;U]  = \langle \psi(x)\bar \psi(y) \rangle   \; ,
\nonumber \\ 
\nonumber \\
\insV &=&  \Gamma_V^k (z,\vec 0) \;,
\nonumber \\ 
\nonumber \\
\insT &=&  \Gamma_T^k (z,\vec 0) \; ,
\end{eqnarray}
that we shall repeatedly use also in the following sections.

\section{Expansion of the vertices}
\label{sec:expvert}
As discussed above, the expression of a correlator at fixed QCD gauge background involves products of traces of propagators $S[U,\lambda^p]$ and operators $\Gamma$, 
as rewritten here below in a very compact notation,
\begin{eqnarray}
C(\vec p) 
&=&  
\langle  C[\vec p;U] \rangle = 
\langle \mbox{Tr}\left\{ S[U,\lambda^p]\, \Gamma_p \cdots \right.  \rangle \; .
\nonumber 
\end{eqnarray}
If the operators are not ultra--local and connect different lattice points, they may also carry a dependence upon $p_k$. For this reason we have used in previous expression the notation $\Gamma_p$. In these situations the calculation of the coefficients of the momentum expansion of the correlators requires the knowledge of the derivatives with respect to external momenta of all $p$-dependent objects, including the vertices, as shown below for the first derivative
\begin{eqnarray}
C^{(1)}_k[U]= 
\mbox{Tr}\left\{ \
\left(\frac{\partial S}{\partial p_k}\, \Gamma \cdots\right)+ 
\left(S \frac{\partial \Gamma}{\partial p_k} \cdots\right)+\ 
\cdots
\right. \; .
\nonumber \\
\end{eqnarray}
One relevant example of $p$-dependent operator is just given by the point--split vector current discussed above. It undergoes the following expansion
\begin{eqnarray}
&&V^k (x,\vec p) = \cos(p_k) V^k(x,\vec 0)-i\sin(p_k)T^k(x,\vec 0) 
\nonumber \\
\nonumber \\
&&\qquad= (1-\frac{p_k^2}{2}+\dots) V^k(x,\vec 0)-i(p_k+\dots) T^k (x,\vec 0) 
\nonumber \\
\end{eqnarray}
to which, in the case of the integrated insertions of these operators, we associate the following graphical representation
\begin{eqnarray}
\sum_z \Gamma_V^k (z,\vec p)&=&\overset{p}{\insV} = \cos(p_k)\insV-i\sin(p_k)\insT 
\nonumber \\ 
\nonumber \\ 
&=& \insV -(i p_k)\insT -\frac{p_k^2}{2}\insV + \dots  \; .
\end{eqnarray}
The previous expressions will be used below in order to expand the integrated correlator proportional to the vacuum polarization tensor.

\section{Applications}
In this section we check the validity of the method by calculating on the lattice the derivatives with respect to external spatial momenta of a set of correlation functions. 

We start by considering two-point correlators of a pseudoscalar density and of an axial current. This example has no relevance from the phenomenological point of view but it is very useful to explain how the method works in practice and to check its effectiveness in a simple and controllable situation. As a second application we consider three-point correlation functions and use our method to extract at zero recoil both form factors parametrizing pseudoscalar--to--pseudoscalar semileptonic transitions. Finally we compute the second derivative of the vacuum polarization tensor and extract the scalar form factor needed for the muon $g-2$ calculation directly at zero momentum. 

The numerical results have been obtained (on a single volume, at fixed lattice spacing and fixed sea quark mass in the $n_f=2$ theory) in order to check the validity of the proposed procedure. For this reason, also in the case of the phenomenologically relevant applications discussed below, we do not quote numbers in physical units neither attempt to estimate possible systematic errors.

For our numerical investigations we have used the gauge configurations corresponding to the $D2$ and $E2$ ensembles of the CLS initiative. These in turn correspond to the simulation parameters $\beta = 5.3$, $k_{sea}=0.13590$, $c_{SW}=1.90952$ and to the lattice volumes $V_{D2}=24^3\times 48$ and $V_{E2}=32^3\times 64$. The larger volume will only be used, in the case of the vacuum polarization tensor, for a test of the method performance with increasing volumes.
All the other information concerning these gauge ensembles can be found in refs.~\cite{DelDebbio:2006cn,DelDebbio:2007pz}.

\subsection{Two--point correlation functions}
A simple and illustrative check of our procedure have been obtained by considering the following two--point correlators
\begin{eqnarray}
&&C_{PP}(t,\vec p)
\nonumber \\
\nonumber \\
&&\quad
=\frac{Z_P^2}{L^6}\langle \sum_{\vec{x},\vec{y}} e^{i\vec p(\vec y-\vec x)} 
[\bar \psi_1 \gamma^5 \psi_2](x)\; [\bar \psi_2 \gamma^5 \psi_1](y)
\rangle
\nonumber \\
\nonumber \\
&&\quad 
=-\frac{Z_P^2}{L^6}\langle \sum_{\vec{x},\vec{y}}
{\text{Tr}\left\{  S[y,x;U] \gamma_5 S[x,y;U,\lambda^p] \gamma_5 \right\}} \rangle \; ,
\nonumber \\
\nonumber \\
\nonumber \\
&&C^\mu_{AP}(t,\vec p)
\nonumber \\
\nonumber \\
&&\quad
=\frac{Z_AZ_P}{L^6}\langle \sum_{\vec{x},\vec{y}} e^{i\vec p(\vec y-\vec x)} 
[\bar \psi_1 \gamma^5 \psi_2](x)\; [\bar \psi_2 \gamma^\mu\gamma^5 \psi_1](y)
\rangle
\nonumber \\
\nonumber \\
&&\quad 
=-\frac{Z_A Z_P}{L^6}\langle \sum_{\vec{x},\vec{y}}
{\text{Tr}\left\{  S[y,x;U] \gamma_5 S[x,y;U,\lambda^p] \gamma^\mu \gamma_5 \right\}} \rangle
 .
\nonumber \\ 
\end{eqnarray}
For large euclidean time separations $t=x_0-y_0$ we can neglect sub--leading exponentials contributing to the correlators and write
\begin{eqnarray}
C_{PP}(t,\vec p) &=&  \frac{G^2}{2E} \; e^{-E t} + \cdots \; ,
\nonumber \\
\nonumber \\
C^\mu_{AP}(t,\vec p) &=&  p^\mu \frac{FG}{2E} \; e^{-E t} + \cdots \; ,
\end{eqnarray}
where $p_\mu$ is the momentum of the pseudoscalar meson, $M$ its mass, $p_0=E=\sqrt{M^2+\vec p^2}$ its energy while $F$  and $G$ are Lorentz-invariant form factors depending upon the meson mass square $p^2=M^2$. 

In this simple situation our method is not needed because the decay constant $F$ of the meson can easily be extracted from the correlators $C_{PP}(t,\vec 0)$ and $C^0_{AP}(t,\vec 0)$ both evaluated at vanishing spatial momentum. On the other hand, we can check our proposal by computing the first coefficient of the momentum expansion of  $C^k_{AP}(t,\vec p)$, 
namely (no sum over $k$)
\begin{eqnarray}
&&\left. \frac{\partial }{\partial p_k} C^k_{A P}(t,\vec p) \right\vert_{\vec p=0}
\nonumber \\
\nonumber \\
&& =
-\frac{Z_AZ_P}{L^6}\langle \sum_{\vec{x},\vec{y}}
{\text{Tr}\left\{  S[y,x;U] \gamma_5 
\frac{\partial }{\partial p_k}S[x,y;U]
\gamma^k \gamma_5 \right\}} \rangle
\nonumber \\
\nonumber \\
&& = i \axialdiagram \; ,
\end{eqnarray}
where, according to the formulae derived in the previous sections, we have
\begin{eqnarray}
\frac{\partial }{\partial p_k}S[x,y;U] &\equiv&
\left.\frac{\partial }{\partial p_k}S[x,y;U,\lambda^p]\right\vert_{\vec p=0}
\nonumber \\
\nonumber \\
&=&-\sum_z{S[x,z;U]\; i\Gamma_V^k(z,\vec 0)\; S[z,y;U]}\;.
\nonumber \\
\end{eqnarray}
We remark that the quantities above have been determined in the periodic theory with $\lambda^p=1$ or, equivalently, $\vec p =0$. The product of propagators appearing in the previous expressions can be conveniently obtained by using the sequential source method, i.e. by using $i\Gamma_V^k(z,\vec 0)\; S[z,y;U]$ as the source vector for a new inversion.

The large time behavior of $\partial C^k_{A P}/\partial p_k$ is now expected to be
\begin{eqnarray}
\left .\frac{\partial }{\partial p_k} C^k_{A P}(t,\vec p)  \right\vert_{\vec p=0}
&=& \frac{FG}{2M} \; e^{-M t} + \cdots \; .
\end{eqnarray}
\begin{figure}[!t]
\includegraphics[width=0.48\textwidth]{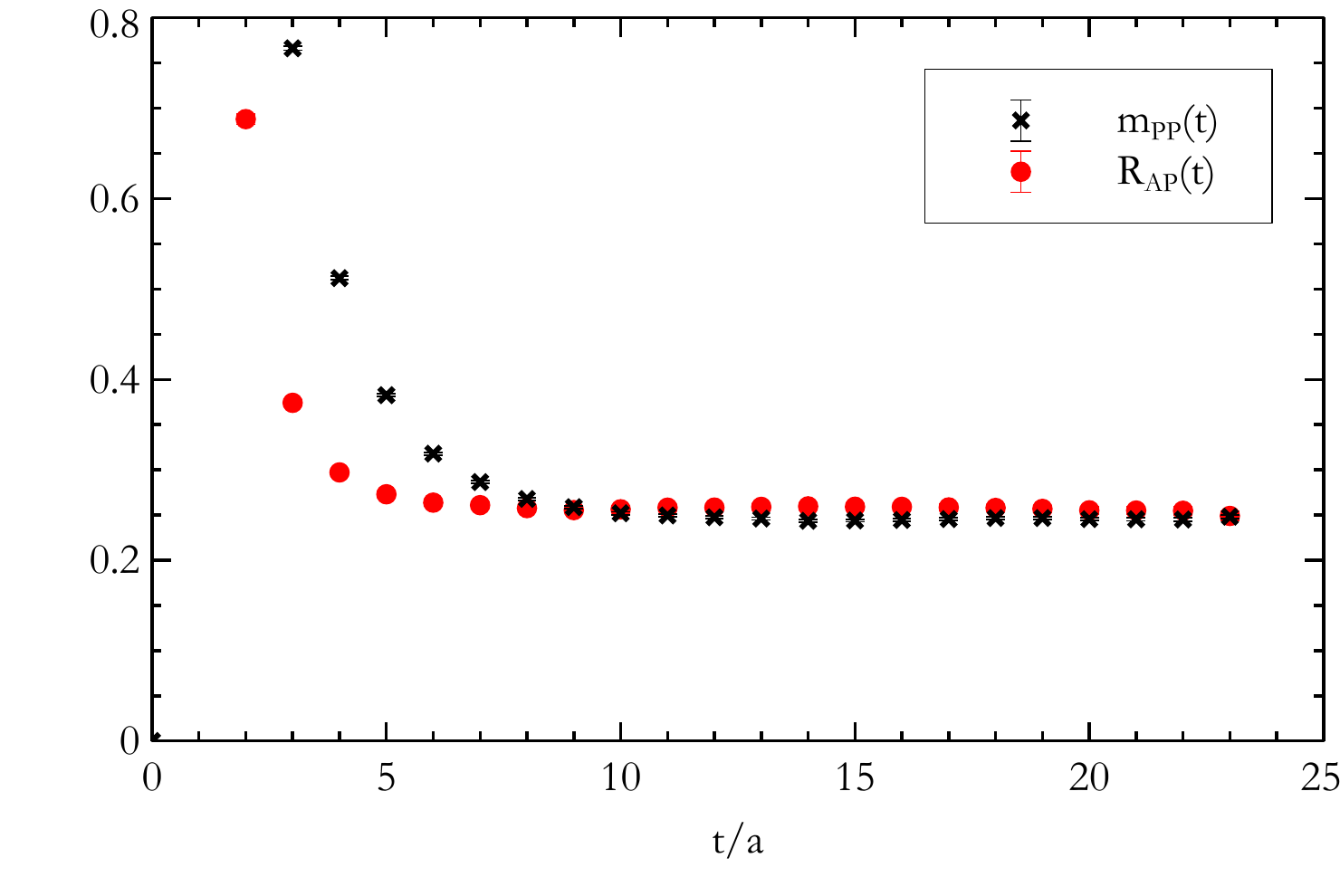}
\caption{\label{fig:axial} 
The black points correspond to the effective mass obtained from the correlation function $C_{PP}(t,\vec 0)$ while the red points correspond to the mass extracted from the ratio $R_{AP}(t)$ (see definition in the text) computed at $\vec p=0$. The two definitions of effective meson mass agree for large euclidean time within statistical errors. The error bars are of the same size of the symbols used for the plots. Data are in lattice units and correspond to the $D2$ gauge ensemble.}
\end{figure}

In order to check this behavior, in FIG.~\ref{fig:axial} we compare the ratio
\begin{eqnarray}
R_{AP}(t)=\frac{C^0_{A P}(t,\vec 0)}{
\left.\frac{\partial }{\partial p_k} C^k_{A P}(t,\vec p) \right\vert_{\vec p=0}}
&=& M + \cdots 
\end{eqnarray}
with the effective mass $m_{PP}(t)$ extracted from the pseudoscalar--pseudoscalar correlator $C_{PP}(t,\vec 0)$ at zero spatial momentum. As it can be seen, for large time separations there is agreement within the statistical errors between the two definitions of effective mass, thus confirming the theoretical expectations and the validity of our method.

An important remark is in order here. Once a given correlator it has been made finite by multiplying it with the appropriate renormalization factors, the derivatives of the renormalized correlation functions with respect to external momenta are also finite, i.e. there is no need of specific renormalization prescriptions for the coefficients of the momentum expansion. In particular, the ratio $R_{AP}(t)$ is finite (renormalization factors identically cancel between the numerator and the denominator) and has indeed been plotted in FIG.~\ref{fig:axial} without multiplying it with any factor.

\subsection{Three--point correlation functions}
In this section we apply our method to three--point correlation functions. 
In particular we consider the form factors entering 
the rate of the semileptonic decay of a pseudoscalar meson into another pseudoscalar meson, $\mathtt{P}_I \rightarrow  \mathtt{P}_F\ell\nu$.
The process is mediated by the vector part of the weak $V-A$ current and the corresponding matrix element can be parametrized in terms of two independent Lorentz-invariant form factors,
\begin{eqnarray}
\langle \mathtt{P}_F\vert  V^\mu  \vert \mathtt{P}_I\rangle =
(p_I+p_F)^\mu  f_+(q^2) + (p_I-p_F)^\mu f_-(q^2) \; ,
\nonumber \\
\end{eqnarray}
where $p_I^\mu$ and $p_F^\mu$ are respectively the initial and final momenta of the hadrons and
\begin{eqnarray}
q^\mu=p_I^\mu-p_F^\mu \; .
\end{eqnarray}
Here and in the following the dependence of form factors upon Lorentz invariants is understood,
$f_\pm=f_\pm(p_I^2,p_F^2,q^2)$ with $p_I^2=M_I^2$ and $p_F^2=M_F^2$.

The form factors are usually extracted by considering ratios of three-point correlators evaluated at different initial and final masses and momenta. By introducing a compact notation for the initial, final and spectator quark propagators ($m_X$ are the quark masses),
\begin{eqnarray}
S(\vec 0) &\mapsto& S[y,x;m_l;U] \; ,
\nonumber \\
\nonumber \\
S^I(\vec p_I) &\mapsto& S[z,y;m_I;U,\lambda^{p_I}] \; ,
\nonumber \\
\nonumber \\
S^F(\vec p_F) &\mapsto& S[x,z;m_F,U,\lambda^{p_F}] \; ,
\end{eqnarray}
these correlators can be written as follows 
\begin{eqnarray}
&&C^\mu_{IF}(t,\vec p_I,\vec p_F)
\nonumber \\
\nonumber \\
&&
=\frac{1}{L^9}\sum_{\vec{x},\vec{y},\vec{z}} 
e^{i\vec p_F(\vec z-\vec x)+i\vec p_I(\vec y-\vec z)}\langle
\nonumber \\
&&\qquad\qquad\quad
[\bar \psi_l \gamma^5 \psi_I](x)\; 
[\bar \psi_F \gamma^\mu \psi_I](z) 
[\bar \psi_I \gamma^5 \psi_l](y)\
\rangle
\nonumber \\
\nonumber \\
&&=
-\frac{1}{L^9}\langle \sum_{\vec{x},\vec{y},\vec{z}}{\text{Tr}\left\{  
S(\vec 0)  \gamma_5 S^F(\vec p_F)  \gamma^\mu S^I(\vec p_I) \gamma_5 
\right\}} \rangle \; .
\nonumber \\
\label{eq:threepointcorr}
\end{eqnarray}
In the previous expression the distance between the interpolating operators of the initial and final pseudoscalar mesons has been set to half of the time extent of the lattice, i.e. $x_0-y_0=T/2$, we have called $t=z_0-y_0$, the improvement of the vector current has been neglected and the multiplicative renormalization factors of the interpolating pseudoscalar densities and of the local vector current have been omitted because these cancel in the ratios from which the form factors are extracted (see below). 

Our gauge ensemble has been generated with two dynamical quarks having the same mass while in eq.~(\ref{eq:threepointcorr}) we are referring to three different quark flavours. The third quark, in the case of different initial and final meson masses, has been treated in the quenched approximation.

When the masses of the initial and final mesons are equal, the conservation of the vector current implies the presence of a single form factor
\begin{equation}
\langle \mathtt{P}_I\vert V^\mu \vert \mathtt{P}_I\rangle =
(p_I+p_F)^\mu \ f_+(q^2)
\end{equation}
exactly normalized to 1 at $q^2=0$. The simple situation of equal initial and final meson masses at zero recoil has no phenomenological relevance but gives us the opportunity of repeating the check already performed in the case of two--point correlators also in the case of three--point correlation functions. To this end, we set the initial spatial momentum to zero, $\vec{p_I}=0$, and consider the behavior of the correlators for large relative time separations by neglecting sub--leading exponentials,
\begin{figure}[!t]
\begin{center}
\includegraphics[width=0.48\textwidth]{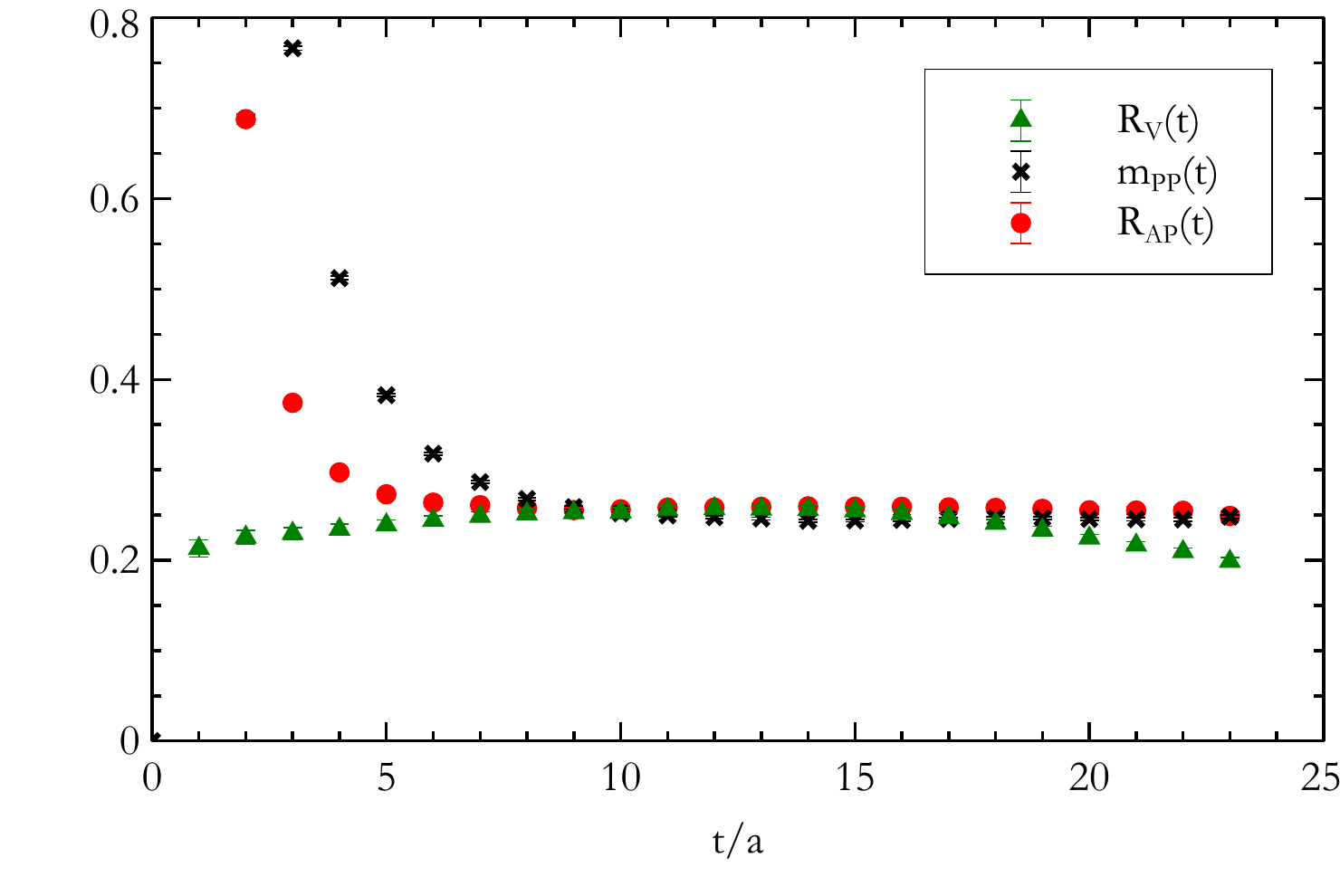}
\caption{\label{fig:RV} 
The black points correspond to the effective mass obtained from the correlation function
$C_{PP}(t,\vec 0)$, the red points correspond to the mass extracted from the ratio $R_{AP}(t)$ while the green points correspond to the mass extracted from the ratio $R_V(t)$ (see definition in the text) computed at $\vec p=0$. The error bars are of the same size of the symbols used to plot the data. Data are in lattice units and correspond to the $D2$ gauge ensemble.
}
\end{center}
\end{figure}
\begin{eqnarray}
C^0_{II}(t,\vec 0,\vec p)&=& (M+E) f_+(q^2) \frac{G^2}{4ME} 
e^{-E (T/2-t) -M t} \; ,
\nonumber \\
\nonumber \\
C^k_{II}(t,\vec 0,\vec p)&=&  p^k f_+(q^2) \frac{G^2}{4ME} 
e^{-E (T/2-t) -M t} \; ,
\nonumber \\
\end{eqnarray}
where $M=M_I=M_F$, $\vec p$ is the spatial momentum of the final meson, $E=\sqrt{M^2+\vec p^2}$ its energy and $G$ the matrix element of the interpolating operator. We have computed the first coefficient of the momentum expansion of $C^k_{II}(t,\vec 0,\vec p)$, i.e. (no sum over $k$)
\begin{eqnarray}
&&\left. \frac{\partial}{\partial p^k} C^k_{II}(t,\vec 0,\vec p)\right\vert_{\vec p=0}
\nonumber \\
\nonumber \\
&&\qquad\quad=
-\frac{1}{L^9}\langle \sum_{\vec{x},\vec{y},\vec{z}}{\text{Tr}\left\{  
S(\vec 0)  \gamma_5 \frac{\partial}{\partial p^k} S^I(\vec 0)  \gamma^k S^I(\vec 0) \gamma_5 
\right\}} \rangle 
\nonumber \\
\nonumber \\
&&\qquad\quad= i \ffq \; ,
\end{eqnarray}
and we have considered the ratio
\begin{eqnarray}
R_V(t)= \frac{1}{2}\frac{C^0_{II}(t,\vec 0,\vec 0)}
{\left. \frac{\partial}{\partial p^k} C^k_{II}(t,\vec 0,\vec p)\right\vert_{\vec p=0}} = M + \cdots \; .
\end{eqnarray}
For large time separations $R_V(t)$ is expected to be equal to the meson mass and, by setting the quark masses equal to the ones used in the previous section, it is possible to compare $R_V(t)$ with $m_{PP}(t)$ and $R_{AP}(t)$. The comparison is shown in FIG.~\ref{fig:RV}. As it can be seen the three definitions of effective meson mass agree within the statistical errors in the middle of the lattice where the large time separation condition is satisfied also for $R_V(t)$.

We now come to the phenomenological relevant situation in which the masses of the two mesons are different. In this case the determination of the two independent form factors can be achieved by solving the following linear system of equations,
\begin{eqnarray}
\langle \mathtt{P}_F\vert V^0 \vert \mathtt{P}_I\rangle &=&
(E_I+E_F) f_+(q^2) + (E_I-E_F)  f_-(q^2)  \; ,
\nonumber \\
\nonumber \\
\langle \mathtt{P}_F\vert V^k \vert \mathtt{P}_I\rangle &=&
(p_I^k+p_F^k) f_+(q^2) + (p_I^k-p_F^k)  f_-(q^2) \; .
\nonumber \\
\end{eqnarray}
The previous system becomes critical at the maximal value of $q^2$, i.e. in the zero recoil situation $q^2_M=(M_I-M_F)^2$. In this case the first equation can be used to obtain a single combination of $f_+$ and $f_-$ that is proportional to $h_+$ defined here below (see ref.~\cite{Hashimoto:1999yp})
\begin{eqnarray}
h_\pm(q^2) = 
\frac{(M_I +M_F)}{2 \sqrt{M_I M_F}} f_\pm(q^2) + \frac{(M_I-M_F)}{2 \sqrt{M_I M_F}} f_\mp(q^2)
\; .
\nonumber \\ 
\end{eqnarray}
The previous parametrization is widely used when the initial and final hadrons are both heavy--light pseudoscalar mesons (for example in the $B\rightarrow D\ell\nu$ case) and the form factors are considered as functions of
\begin{eqnarray}
\omega=\frac{p_I\cdot p_F}{M_IM_F} \; .
\end{eqnarray}
The zero recoil point at $q^2_M$ corresponds to $\omega=1$.

By applying our method it is possible to obtain both the form factors directly at $q^2_M$. To this end it is useful to consider (by neglecting sub--leading exponentials)
\begin{eqnarray}
&&\left. \frac{\partial}{\partial p^k} 
C^k_{IF}(t,\vec p,\vec 0)\right\vert_{\vec p=0} = i\ifp
\nonumber \\
\nonumber \\
&&\qquad=
[f_+ +f_-](q^2_M) \frac{G_IG_F}{4M_IM_F} e^{-M_F (T/2-t) -M_I t} \; ,
\nonumber \\
\nonumber \\
\nonumber \\
&&\left. \frac{\partial}{\partial p^k} 
C^k_{IF}(t,\vec 0,\vec p)\right\vert_{\vec p=0} = i\ifq
\nonumber \\
\nonumber \\
&&\qquad=
[f_+ -f_-](q^2_M) \frac{G_IG_F}{4M_IM_F} e^{-M_F (T/2-t) -M_I t} \; .
\nonumber \\
\end{eqnarray}
In previous expressions one of the quark lines has been drawn with a different color (green) in order to represent a quark of different mass with respect to the others (in particular in the numerical analysis we have set $k_{green}=0.13620$ to be compared with $k_{sea}=0.13590$).
By relying on the hermiticity of the vector current we also get
\begin{eqnarray}
&&\left. \frac{\partial}{\partial p^k} 
C^k_{FI}(t,\vec 0,\vec p)\right\vert_{\vec p=0} = i\fiq
\nonumber \\
\nonumber \\
&&\qquad=
[f_+ +f_-](q^2_M) \frac{G_IG_F}{4M_IM_F} e^{-M_I (T/2-t) -M_F t} \; ,
\nonumber \\
\nonumber \\
\nonumber \\
&&\left. \frac{\partial}{\partial p^k} 
C^k_{FI}(t,\vec p,\vec 0)\right\vert_{\vec p=0} = i\fip
\nonumber \\
\nonumber \\
&&\qquad=
[f_+ -f_-](q^2_M) \frac{G_IG_F}{4M_IM_F} e^{-M_I (T/2-t) -M_F t} \; .
\nonumber \\
\end{eqnarray}
By using these relations, together with the corresponding ones in the case of coinciding initial and final external states we get
\begin{eqnarray}
&&N_{II}(t)=
\sqrt{-\iip \iiq}=
\frac{G_I^2}{4M_I^2} e^{-M_I T/2} ,
\nonumber \\
\nonumber \\
&&N_{FF}(t)=
\sqrt{-\ffp \ffq}=
\frac{G_F^2}{4M_F^2} e^{-M_F T/2} ,
\nonumber \\
\nonumber \\
&&R_+(t)=\sqrt{\frac{-\ifp\fiq}{N_{II}(t)N_{FF}(t)}}=[f_++ f_-](q^2_M) ,
\nonumber \\
\nonumber \\
&&R_-(t)=\sqrt{\frac{-\ifq\fip}{N_{II}(t)N_{FF}(t)}}=[f_+ - f_-](q^2_M) .
\nonumber \\
\end{eqnarray}
\begin{figure}[!t]
\begin{center}
\includegraphics[width=0.48\textwidth]{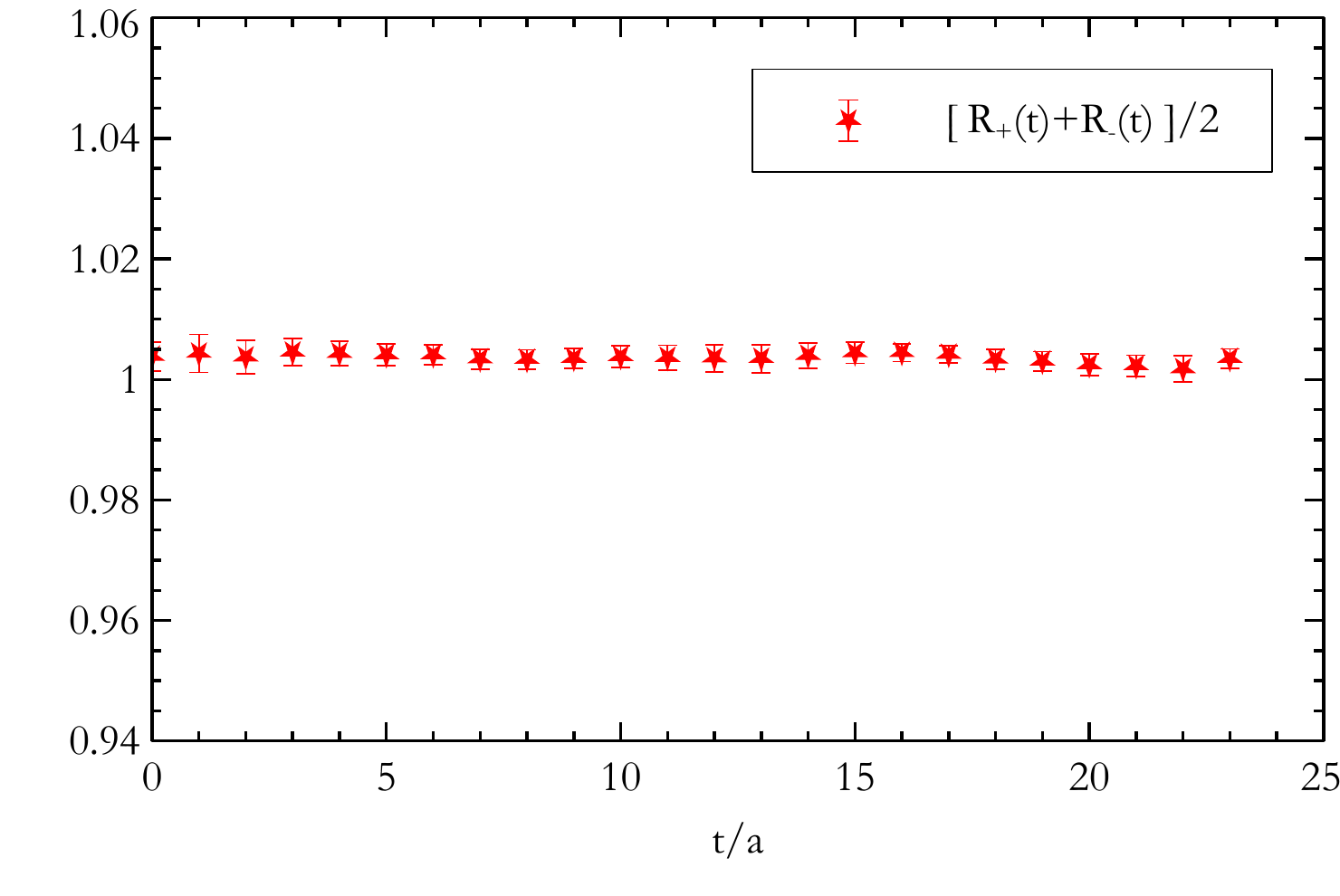}
\includegraphics[width=0.48\textwidth]{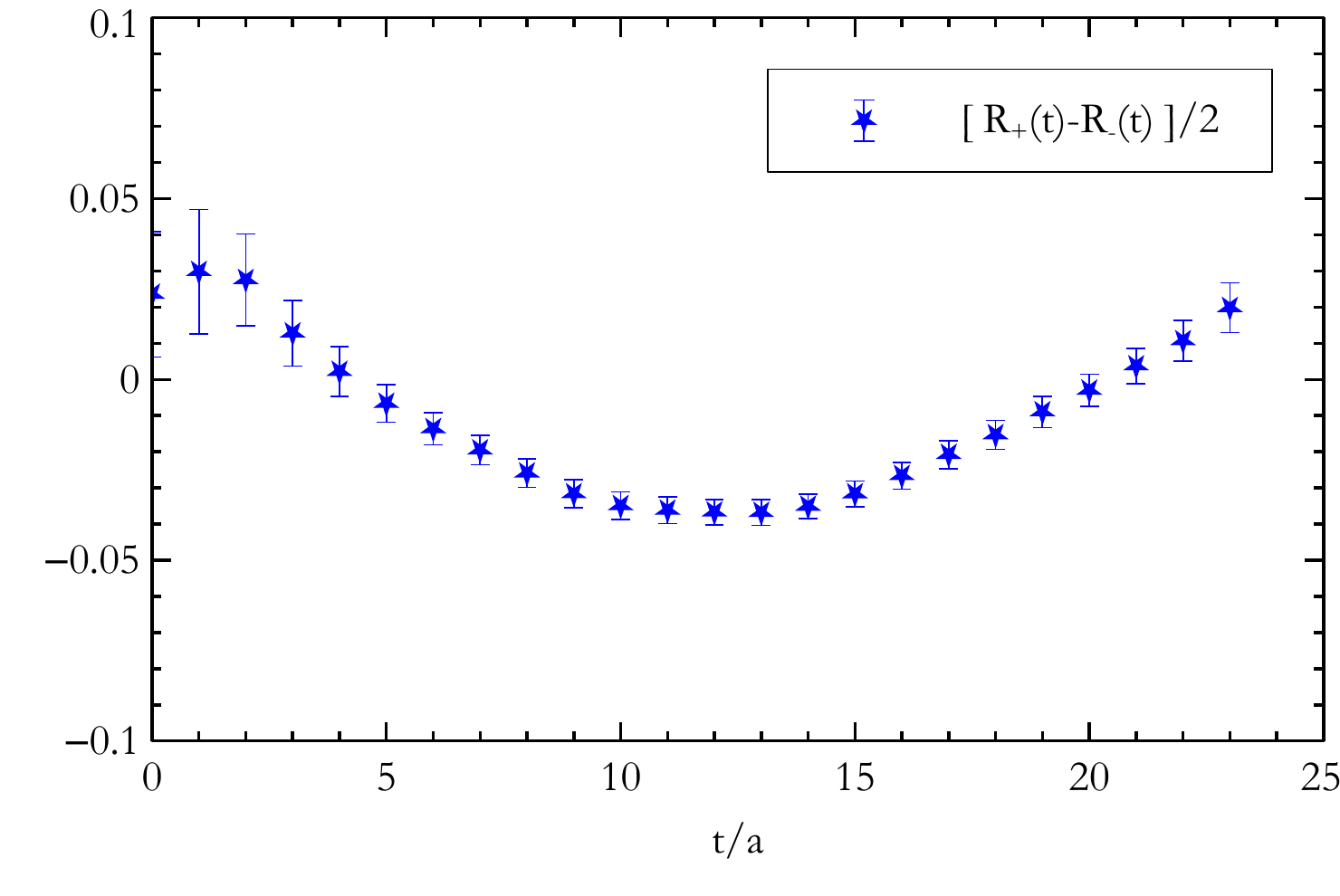}
\caption{\label{fig:fpfm} 
In the top panel we plot the effective form factor $f_+(q^2_M)$ while in the
bottom panel we show the effective form factor $f_-(q^2_M)$ for which we get a good plateaux in the middle of the lattice where the large time separation condition is satisfied. Data correspond to the $D2$ gauge ensemble.
}
\end{center}
\end{figure}
The effective form factors are then obtained from
\begin{eqnarray}
&&f_+(q^2_M)=\frac{R_+(t)+R_-(t)}{2} \; ,
\nonumber \\
\nonumber \\
&&f_-(q^2_M)=\frac{R_+(t)-R_-(t)}{2} \; .
\end{eqnarray}
The combinations of correlators given in previous relations is particularly effective in minimizing statistical fluctuations that cancel between the different factors. 
In particular, when the initial and final meson masses are equal, the normalization conditions $f_+(0)=1$ and $f_-(0)=0$ are identically verified for each gauge configuration of the given ensemble. To give an idea of the plateaux that can be obtained by using this method we show in FIG.~\ref{fig:fpfm} the effective form factors extracted from our data.

\subsection{The vacuum polarization tensor}
In this section we show how the momentum expansion can profitably be used in order to calculate the hadronic vacuum polarization tensor, a quantity which is needed in order to predict the leading hadronic contribution to the anomalous magnetic moment of the muon.

The hadronic vacuum polarization tensor is calculated on the lattice by considering the integrated two--point correlator of the electromagnetic currents of the quarks
\begin{eqnarray}
C^{\mu\nu}(p)&=&
\frac{1}{(TL^3)^2}\sum_{x,y}e^{ip(y-x+\hat{\nu}/2-\hat{\mu}/2)} \langle 
V^\mu_{em}(x)\ V^\nu_{em}(y)
\rangle
\nonumber \\
\nonumber \\
&=&(\delta^{\mu\nu}\hat{p}^2-\hat{p}^\mu \hat{p}^\nu)\Pi(p^2) \; ,
\label{eq:latcmunu}
\end{eqnarray}
where $\hat{p}^\mu=2\sin(p^\mu/2)$. Given our choice of lattice Dirac operator, we have used the point--split vector currents in order to define the electromagnetic currents of the quarks,
\begin{eqnarray}
V^\mu_{em}(x) \rightarrow [\bar \psi \Gamma^\mu_V \psi](x,\vec 0) \; .
\end{eqnarray}
We have chosen the momentum routing in which the external momentum $p$ flows through one of the two quark propagators. With this choice the point--split vector currents at the vertices connect two fermion lines with different momentum and, consequently, bring the dependence upon $\vec p/2$. 

The leading hadronic contribution to the $g-2$ of the muon is obtained 
by extracting the scalar form factor $\Pi(p^2)$ from eq.~(\ref{eq:latcmunu}) and by considering
the difference $\Pi(p^2)-\Pi(0)$. The subtracted form factor is introduced in order to cancel divergent contributions (see for example refs.~\cite{Gockeler:2003cw,Aubin:2006xv,DellaMorte:2011aa,Feng:2011zk,Boyle:2011hu} for dedicated works) and, for this reason, it is very important to have an accurate determination of $\Pi(0)$. By using standard techniques, $\Pi(0)$  can only be obtained by extrapolating the data obtained at $p^2>0$. These extrapolations unavoidably introduce systematic errors (for a recent discussion of this point see~\cite{Aubin:2012me}). In the following we shall show how $\Pi(0)$ can be computed on the lattice directly and without the need of any extrapolation. 

As for the other quantities computed in this paper, the results for $\Pi(p^2)$ have been obtained with limited statistics, on a single volume, at fixed quark masses, etc. in order to demonstrate the effectiveness of the proposed procedure. For this reason we shall not use our results to make a prediction for the muon $g-2$. This will be the subject of future work.

After fermion integration and Wick contractions, the correlator $C^{\mu\nu}(p)$ has both fermion--connected and fermion--disconnected contributions. If, for example, we limit ourself to the $N_f=2$ case in the isospin symmetric theory we have
\begin{eqnarray}
C^{\mu\nu}(p)= 
(e_u +e_d)^2\disc -(e_u^2 +e_d^2)\conn.
\end{eqnarray}
where $e_u=2/3$ and $e_d=-1/3$.
In the following we shall concentrate on the connected part. 
Actually, many of the lattice collaborations involved in the computation of the hadronic vacuum polarization tensor~\cite{Gockeler:2003cw,Aubin:2006xv,DellaMorte:2011aa,Feng:2011zk,Boyle:2011hu} have neglected (or just attempted an estimate of) disconnected contributions. The choice is due to the big computational effort required for the calculation and not because these are expected (at least in general) to be negligible. Our method has to be generalized for disconnected diagrams. 

In the following we call $\hat C^{\mu\nu}$ the connected contribution corresponding to a single quark without multiplying it for the corresponding electric charge. In particular, we consider the correlators with different spatial indices  ($\mu=k, \nu=h\neq k$) because these are  proportional to the spatial momenta and because they do not require additional contributions in order to satisfy gauge Ward identities. 
Indeed, the lattice photon self energy with equal indices includes also the tadpole graph, where two photons couple to the fermion line in a single vertex 
(the current coupled to two photons is the one that we have previously indicated as ``tadpole current''). We checked anyway, on limited statistics, that the gauge Ward identities $\sum_\mu \hat{p}_\mu \hat C^{\mu\nu}=\sum_\nu\hat{p}_\nu \hat C^{\mu\nu}(p)=0$ are satisfied.

First, fixed $\mu=1$ and $\nu=2$, we have computed the integrated correlation at $p^1>0$ and $p^2>0$ and divided it by the momenta,
\begin{eqnarray}
&&\Pi(p^2>0)=-\frac{\hat C^{12}(p)}{\hat{p}^1 \hat{p}^2}=\frac{1}{(TL^3)^2}\sum_{x,y} 
\langle 
\nonumber \\
\nonumber \\
&& \quad
{\text{Tr}\left\{  S[y,x;U]  \Gamma_V^1(x,\vec p/2) 
S[x,y;U,\lambda^p] \Gamma_V^2(y,\vec p/2) \right\}} 
\rangle  .
\nonumber \\
\label{eq:pino0}
\end{eqnarray}
Then we have applied the rules discussed in the previous sections to define  the second mixed derivative, acting on propagators and vertices and evaluated at zero momentum, according to
\begin{eqnarray}
&&\Pi(0) =
-\left.\frac{\partial^2 \hat C^{12}(p)}{\partial p_1 \partial p_2}\right |_{p^2=0} = 
\frac{1}{(TL^3)^2}\sum_{x,y}
\langle 
\nonumber \\
\nonumber \\
\nonumber \\
&&\qquad
\text{Tr}\left[  S  \Gamma_V^1 \frac {\partial^2 S}{\partial p_1\partial p_2} \Gamma_V^2 \right] 
-\frac{1}{4} \text{Tr}\left[  S  \Gamma_T^1 S  \Gamma_T^2 \right] 
\nonumber \\
\nonumber \\
&& \qquad 
-\frac{i}{2}\text{Tr}\left[  S  \Gamma_T^1\frac{\partial S}{\partial p_2}  \Gamma_V^2 \right] 
-\frac{i}{2}\text{Tr}\left[  S  \Gamma_V^1 \frac {\partial S}{\partial p_1} \Gamma_T^2 \right]
\rangle \; ,
\nonumber \\
\label{eq:pi0}
\end{eqnarray}
where, for sake of brevity, we have dropped position arguments and we have used the relations
\begin{eqnarray}
\frac{\partial \Gamma_V^k(x,\vec p/2)}{\partial p_k} = -\frac{i}{2} \Gamma_T^k(x,\vec p/2) \; ,
\nonumber \\
\nonumber \\
\frac{\partial \Gamma_T^k(x,\vec p/2)}{\partial p_k} = -\frac{i}{2} \Gamma_V^k(x,\vec p/2) \; ,
\end{eqnarray}
to obtain the derivative of the vertices (see section~\ref{sec:expvert} above). Note that in the previous expressions the factor $1/2$ appears because the currents here depend upon $\vec p/2$ and not upon $\vec p$. 

\begin{figure}[!t]
\begin{center}
\includegraphics[width=0.48\textwidth]{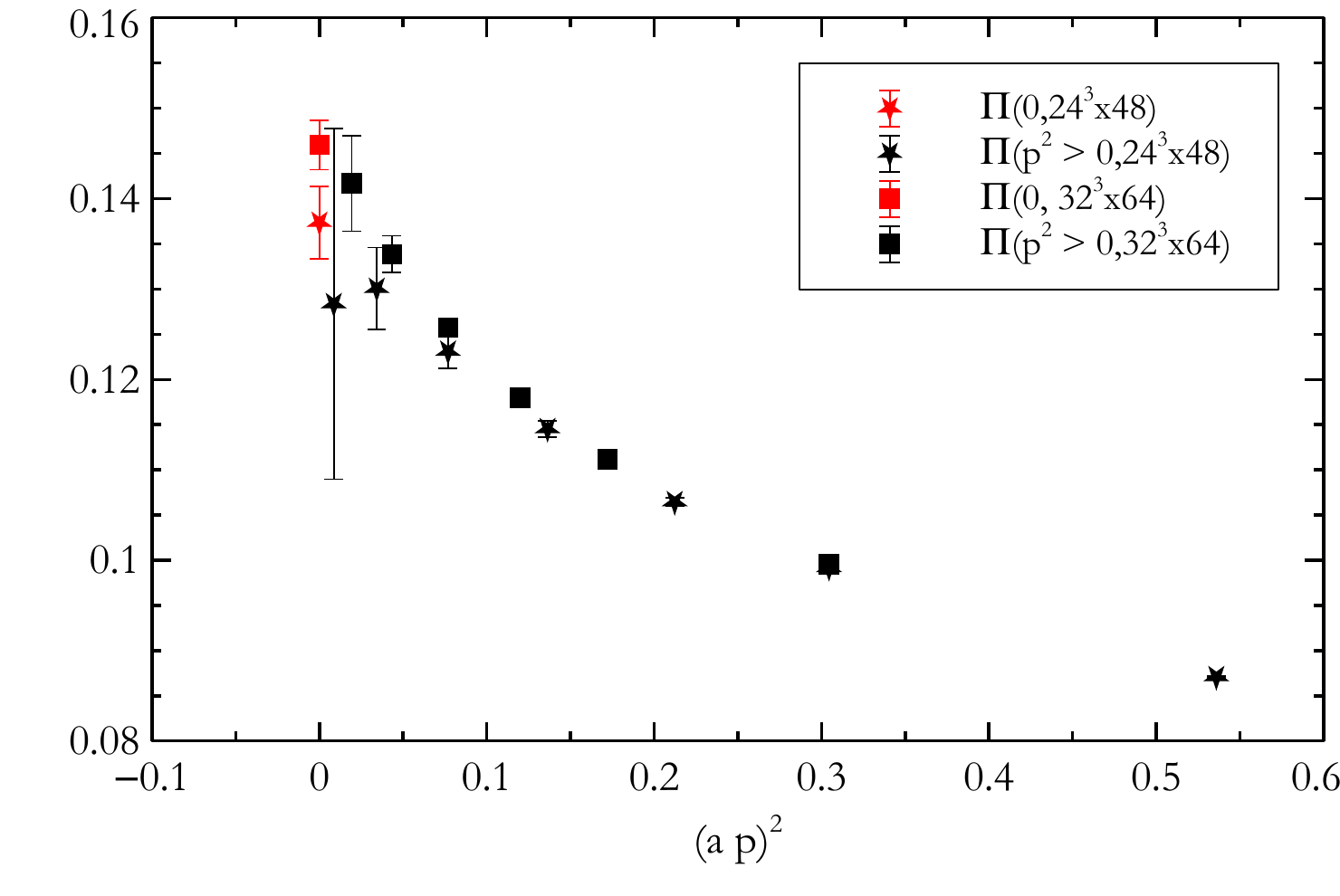}
\caption{\label{fig:VP} 
Black points correspond to the calculation of $\Pi(p^2)$ performed by using standard techniques (according to eq.~(\ref{eq:pino0})) on two lattice volumes, $24^3\times 48$ for the $D2$ ensemble and $32^3\times 64$ for the $E2$ ensemble. The red points correspond to $\Pi(0)$ calculated directly on the lattice (according to eq.~(\ref{eq:pi0})) for the two volumes. Data are in lattice units.
}
\end{center}
\end{figure}
To the lattice definition of $\Pi(0)$, eq.~(\ref{eq:pi0}), it can be given the following graphical representation (see also eq.~(\ref{eq:graphvertexp}))
\begin{eqnarray}
\Pi(0) &=&
-\left.\frac{\partial^2 \hat C^{12}(p)}{\partial p_1 \partial p_2}\right |_{p^2=0} 
\nonumber \\
\nonumber \\
&=&
-\Pcone -\Pctwo
\nonumber \\
\nonumber \\
&&- \frac{1}{2} \Pa -\frac{1}{2} \Pd-\frac{1}{4} \Pb
 \; .
\nonumber \\
\end{eqnarray}

In FIG.~\ref{fig:VP} we show our results. 
The black points correspond to $\Pi(p^2)$ obtained from eq.~(\ref{eq:pino0}) and, as expected, tend to be noisy for small values of $p^2$. The red points correspond to $\Pi(0)$ calculated directly on the lattice according to eq.~(\ref{eq:pi0}). The data, obtained with limited statistics ($150$ gauge configurations for the $D2$ ensemble and $138$ gauge configurations for the $E2$ ensemble), correspond to two different lattice volumes ($V_{D2}=24^3\times 48$ and $V_{E2}=32^3\times 64$) and differ at small momenta for finite volume effects.

For each data set, the error on $\Pi(0)$ is comparable to the error that can be obtained at $(ap)^2\sim 0.05$ but, coming from a direct calculation, it does not need to be corrected for systematic errors due to extrapolations and, important to note, it scales with the statistics.
Furthermore, the error on $\Pi(0)$ scales favorably with the lattice volume.

\section{Conclusions}

The method discussed in this letter allows the direct calculation on the lattice of the derivatives of correlators with respect to external momenta. We have described the method and
checked its validity for several correlation functions.

In particular, we have derived expressions to be used in order to compute both form factors parametrizing semileptonic decays of pseudoscalar mesons into other pseudoscalar mesons, directly at zero recoil. These relations, checked numerically in this paper, may have many important phenomenological applications, for example in the calculation of $B\rightarrow D\ell\nu$ differential decay rate without excluding the $\ell=\tau$ case, etc. Similar relations can be easily derived along the lines discussed in this paper for the pseudoscalar-to-vector channels.

Another important application of the method concerns the direct calculation of the scalar form factor $\Pi(p^2)$ needed to predict the hadronic vacuum polarization contribution to the muon anomalous magnetic moment. In particular our method, as explicitly shown in the paper, allows to obtain $\Pi(0)$ without the need of any extrapolation and related systematic uncertainties.

The method is general, easy to implement, and we are pretty confident it will find several applications in the field of lattice phenomenology.

\begin{acknowledgments}
We thank F.~Sanfilippo for useful discussions,
our colleagues of the CLS initiative for sharing the efforts of the generation of the gauge configurations used in this work and MIUR for partial support under the contract PRIN09.
\end{acknowledgments}


\end{document}